\documentclass[11pt,a4paper,english,nofootinbib,,superscriptaddress]{revtex4}
\usepackage{lmodern}

\usepackage[T1]{fontenc}
\usepackage[latin9]{inputenc}
\usepackage{babel}

\usepackage{amsmath}
\usepackage{graphicx}
\usepackage{amssymb}
\usepackage{esint}
\usepackage[unicode=true, pdfusetitle,
 bookmarks=true,bookmarksnumbered=false,bookmarksopen=false,
 breaklinks=false,pdfborder={0 0 1},backref=false,colorlinks=false]
 {hyperref}
\def\b{\begin{equation}}
\def\e{\end{equation}}

\makeatletter
\@ifundefined{textcolor}{}
{%
 \definecolor{BLACK}{gray}{0}
 \definecolor{WHITE}{gray}{1}
 \definecolor{RED}{rgb}{1,0,0}
 \definecolor{GREEN}{rgb}{0,1,0}
 \definecolor{BLUE}{rgb}{0,0,1}
 \definecolor{CYAN}{cmyk}{1,0,0,0}
 \definecolor{MAGENTA}{cmyk}{0,1,0,0}
 \definecolor{YELLOW}{cmyk}{0,0,1,0}
 }

\usepackage{latexsym}\usepackage{bm}

\makeatother

\makeatother

\begin{document}

\title{Unitarity of Weyl-Invariant New Massive Gravity and Generation of Graviton Mass via Symmetry Breaking}

\author{M. Reza Tanhayi}

\email{m_tanhayi@iauctb.ac.ir}

\affiliation{Department of Physics,\\
 Middle East Technical University, 06531, Ankara, Turkey}

\affiliation{Department of Physics,\\
Islamic Azad University Central Tehran Branch, Tehran, Iran}

\author{Suat Dengiz}

\email{suat.dengiz@metu.edu.tr}

\affiliation{Department of Physics,\\
 Middle East Technical University, 06531, Ankara, Turkey}

\author{Bayram Tekin}

\email{btekin@metu.edu.tr}

\affiliation{Department of Physics,\\
 Middle East Technical University, 06531, Ankara, Turkey}

\date{\today}
\begin{abstract}

We give a detailed analysis of the particle spectrum and the
perturbative unitarity of the recently introduced Weyl-invariant
version of the new massive gravity in 2+1 dimensions. By computing
the action up to second order in the fluctuations of the metric,
the gauge and the scalar fields around the anti-de Sitter (AdS)
and flat vacua, we find that the theory describes unitary (tachyon
and ghost-free) massive spin-2, massive (or massless) spin-1 and
massless spin-0 excitations for certain ranges of the
dimensionless parameters. The theory is not unitary in de Sitter
space. Scale invariance is either broken spontaneously (in AdS
background) or radiatively (in flat background) and hence the
masses of the particles are generated either spontaneously or at
the second loop order.

\end{abstract}
\maketitle

\section{Introduction}

Einstein's general relativity (GR) is expected to be modified at
both large (astrophysical) (IR) and small (UV) regions. There are
ample theoretical (in the case of UV) and experimental (in the
case of IR) reasons to conclude that GR can only be an effective
theory that works perfectly in the intermediate regions, such as
the solar system and \emph{etc}. Apriori, the nature of UV and IR
modifications is quite different. For UV modifications, experience
from quantum field theory dictates that if one is to define a
perturbatively well-behaved (that is renormalizable and unitary)
gravity theory, then one must introduce higher powers of curvature
that modify both the tree-level propagator structure and the
interactions. Unfortunately, it is well-known that such a theory
simply does not exist in four dimensions \cite{stelle}. On the
other extreme, IR modifications consist of introducing a
cosmological constant and/or mass to the graviton. Even though,
theoretically, cosmological Einstein theory is the easiest
extension of GR, the problems with the cosmological constant are
well-known (such as the difficulty of keeping it small in the
quantum theory). Graviton mass on the other hand is a very subtle
issue. Given a massless free spin-2 field about a maximally
symmetric background, one can introduce the Fierz-Pauli term
respecting the background symmetries to get a massive spin-2
field. But such a theory does not seem to arise from a
diffeomorphism invariant interacting gravity theory save the
unique case of the 2+1 dimensions.

For $ D= 2+1 $ dimensions, new massive gravity (NMG) introduced in
\cite{BHT} provides a non-linear extension of the
Fierz-Pauli massive spin-2 theory. For the mostly plus signature
the action reads\footnote{To have a maximally symmetric vacuum one
must have $\lambda$ > -1 and one can normalize $\sigma^2=1$ in
NMG. On the other hand, $\lambda =0$ and  $\sigma$ should be free
in the Weyl-invariant version, since the numerical values of
various couplings play a key role in the unitarity analysis.}
\begin{equation}
 I_{NMG}=\frac{1}{\kappa^2} \int d^3 x \sqrt{-g} \Big [\sigma R-2 \lambda  m^2 + \frac{1}{m^2} \Big ( R^2_{\mu\nu}-\frac{3}{8}R^2 \Big )\Big],
\label{nmg}
\end{equation}
which attracted a lot of attention: Detailed works on it appeared
in
\cite{BHT,GulluTekin,deser,nakasone,liusun,canonical,cubic,Aliev}
regarding its unitarity, solutions and \emph{etc}. Remarkably,
higher curvature terms in this theory provide in some sense both
the viable UV and IR modifications that one is interested in.
Unfortunately, this state of happy affairs do not extend to four
dimensions. But in any case, 2+1 dimensional gravity is a valuable
theoretical lab for ideas in quantum gravity. [In fact, according
to the proposal of Horava for which spacetime's spectral dimension
reduces at high energies, 3D gravity becomes much more relevant
and NMG appears as part of the non-covariant 3+1 dimensional
action \cite{horava}.]

Having understood that NMG describes a consistent
parity-invariant\footnote{ Parity non-invariant massive spin-2
theory with a single helicity degree of freedom, that is the
Topologically Massive Gravity, was found in 1982 \cite{DJT}.}
massive spin-2 theory in 2+1 dimensions, the next natural question
is to ask if graviton mass can be generated from breaking a
symmetry (not diffeomorphism invariance) in this theory in analogy
with the Higgs mechanism in the Standard Model. This question was
answered in the affirmative recently in \cite{DengizTekin} by
finding a local scale-invariant (Weyl-invariant) version of NMG
and showing that in the case of (A)dS space, the vacuum breaks the
conformal symmetry spontaneously and for flat space conformal
symmetry is broken at the two loop level \cite{tantekin} via the
Coleman-Weinberg mechanism \cite{coleman}. Referring to
\cite{DengizTekin} for the details of how the Weyl-invariant
extension of NMG (and other higher curvature models, such as the
Born-Infeld NMG \cite{gullusismantekin}) was introduced and how
symmetry gets broken, here we just quote the final expression
\begin{equation}
\begin{aligned}
  S_{WNMG}&= \int d^3 x \sqrt{-g} \bigg \{\sigma \Phi^2 \Big(R-4\nabla \cdot A -2A^2  \Big) \\
& \qquad \quad \qquad \quad +\Phi^{-2} \bigg [R^2_{\mu\nu}-\frac{3}{8} R^2-2 R^{\mu\nu}\nabla_\mu A_\nu+ 2R^{\mu\nu}A_\mu A_\nu \\
& \qquad \quad \qquad \quad\qquad \quad   +R\, \nabla \cdot A-\frac{1}{2}R A^2+2  F_{\mu\nu}^2 +(\nabla_\mu A_\nu)^2 \\
& \qquad \quad\qquad \quad \qquad \quad -2 A_\mu A_\nu \nabla^\mu
A^\nu-(\nabla \cdot A)^2+\frac{1}{2}A^4 \bigg ] \bigg
\}+S_\Phi+S_{A_\mu},
 \label{winmg}
\end{aligned}
\end{equation}
where $S_\Phi$ and $S_{A_\mu}$ are the Weyl-invariant scalar and
gauge field actions, which are given by
\begin{equation}
\begin{aligned}
& S_{\Phi}=- \frac{1}{2}\int d^3 x \sqrt{-g}\bigg \{\Big (
\partial_\mu \Phi-\frac{1}{2} A_\mu\Phi\Big)^2+\nu\Phi^6\bigg\},\\
&S_{A_\mu}= \beta \int d^3 x \sqrt{-g}\,\, \Phi^{-2}
F^2_{\mu \nu}.
\end{aligned}
\end{equation}
The action (\ref{winmg}) is invariant under
the simultaneous transformation of the metric and the fields as
\begin{equation}\label{gaugetran}
 g_{\mu\nu} \rightarrow g^{'}_{\mu\nu}=e^{2 \zeta(x)} g_{\mu\nu}, \hskip 1 cm \Phi \rightarrow \Phi^{'} =e^{-\frac{(n-2)}{2}\zeta(x)}
 \Phi, \hskip 1 cm A_\mu \rightarrow A^{'}_\mu = A_\mu - \partial_\mu \zeta(x).
\end{equation}
It is important to note that there are no dimensionful parameters
in the theory, on the other hand, local scale invariance does not
fix the relative numerical coefficients of various parts which are
independently scale-invariant. Generically, up to a numerical
scaling of the total action, there are 4 dimensionless parameters
that one can introduce.  By scaling the total action, we set  the
coefficient of the kinetic part of the scalar action to its
canonical non-ghost form and to keep contact with the NMG, we take
the numerical coefficient of the quadratic part of the action to
be 1 (this choice can easily be relaxed). Therefore, we have 3
dimensionless parameters $\sigma, \nu, \beta$.

The Weyl-invariant theory (\ref{winmg}) obviously is much larger
than NMG (\ref{nmg}) in the sense that when one sets $
\Phi=\sqrt{m} $, $ \nu=2 \lambda $ and $ A_\mu= 0 $, at the level
of the action, one recovers NMG (\ref{nmg}), with a fixed
gravitational coupling $ \kappa=m^{-1/2} $ and a fixed
cosmological. [In fact all the dimensionful scales are
determined by the symmetry breaking order parameter $ <\Phi>
=\Phi=\sqrt{m} $]. Let us consider the infinite dimensional field
space $ {\cal M}=[g_{\mu \nu}, A_\mu, \Phi]$ to be the space of
all fields satisfying the field equations derived from the action
(\ref{winmg}). It was shown in  \cite{DengizTekin} that
"NMG-point", that is [$ g_{\mu \nu}, 0, \sqrt{m} $], is in ${\cal
M}$. Moreover, if one freezes the scalar and the gauge fields to
these NMG-point values and consider fluctuations just in the
metric directions, one exactly gets the same spectrum as NMG
around its AdS or flat vacua. This was shown in \cite{DengizTekin}
but what was left out in that work and which will be remedied
here, is a complete study of the second order fluctuations of all
the fields around the NMG point (or the vacuum of the theory).
Namely, apriori, the stability and unitarity of (\ref{winmg}) is
not clear for all allowed fluctuations in the metric, gauge and
scalar field directions on  ${\cal M}$. The main task of this
paper is to show that NMG-point is  stable by proving that there
are no ghosts and tachyons in the particle spectrum of the
Weyl-invariant action (\ref{winmg}). Therefore, mass of the
graviton and the mass of the gauge field is consistently generated
by the symmetry breaking mechanism of the conformal symmetry.

The layout of the paper is as follows: In section II, we first find the expansion of the action up to second order
in the fields around the (A)dS or flat vacua. This section also discusses issues about Jordan versus Einstein frame
and the Weyl-invariant gauge-fixing in the gauge sector. In section III, we decouple the fields and identify
the masses and also the unitarity regions of the dimensionless parameters. We collect some useful computations in the appendices.

\section{Quadratic fluctuations about the vacuum}

In \cite{DengizTekin}, the field equations coming from the action
(\ref{winmg}) and its vacuum solution were given. Here, we do not
depict the field equations, since they are rather lengthy, instead
we note that the vacuum solution (let us first take a dS or AdS
vacuum, as the flat vacuum will follow these) is given as
\begin{equation}
 \Phi_{vac}= \sqrt{m}, \hskip 1 cm A^\mu_{vac}=0, \hskip 1 cm g_{\mu \nu}=\bar{g}_{\mu \nu},
\end{equation}
here $ \bar{R}_{\mu \nu}=2 \Lambda \bar{g}_{\mu \nu} $. And the
cosmological constant satisfies\footnote{ Here we take the point
of view that $\Phi_{vac}$ is given and $\Lambda$ is determined,
one can also take a different point of view that $\Lambda$ is
given and $\Phi_{vac}$ is determined. See \cite{DengizTekin} for a
discussion on this.}
\begin{equation}
\Lambda^2+4 \sigma m^2 -\nu m^4 =0.
\label{vacuum}
\end{equation}
Given $m^2$, generically, there are two vacua
\begin{equation}
\Lambda_{\pm}= m^2 \Big[ - 2 \sigma \pm \sqrt{4\sigma^2 + \nu} \Big ].
\label{lambda_exp}
\end{equation}
Our task now is to study the stability of these vacua and also
study the particle spectrum of the model. This can be achieved by
considering the second order fluctuations about the vacuum following from
\begin{equation}
 \Phi=\sqrt{m}+\tau \Phi_{L}, \hskip 1 cm   A_\mu=\tau A^{L}_\mu, \hskip 1 cm g_{\mu \nu}=\bar{g}_{\mu \nu}+\tau h_{\mu \nu},
\label{agg}
\end{equation}
where we have introduced $ \tau$, a small dimensionless parameter
to keep the track of the expansion orders. In what follows, we
will use the conventions given in \cite{ubinmg}. The
expansions of various curvature terms are needed in the
computations, so we collect them in  Appendix A.

Since the action (\ref{winmg}) is highly complicated with fields
coupled to each other, it is a non-trivial task to find the basic
oscillators (free particles) of the theory. There are a couple of
paths one can take. For example, one can linearize field equations
and try to decouple the fields. Or, one can transform the action
to the Einstein frame and then find the field equations and do the
linearization. These two paths do not give an efficient way for
the study of the spectrum. [See Appendix B for the Einstein frame version of the Weyl-invariant quadratic theory.]
 As a third way, one can directly
compute the action up to quadratic order in the fluctuations about
its vacua, which we shall adopt here. This will lead to coupled
fields at the quadratic level. Then we will find a way to decouple
the basic free fields in the theory. This procedure is quite
lengthy but there seems to be no way of avoiding it and it is
still easier than the above mentioned procedures. The action
(\ref{winmg}), after making use of the field fluctuations
(\ref{agg}) and the relevant formulas in the Appendix A, can be
written as
\begin{equation}
 S_{WNMG}=\bar{S}_{WNMG}+\tau S^{(1)}_{WNMG}+\tau^2 S^{(2)}_{WNMG}+{\cal
 O}(\tau^3),
\end{equation}
where $ \bar{S}_{WNMG} $ is the value of action evaluated in the
background which is irrelevant for our purposes. On the other hand
$ S^{(1)}_{WNMG} $ vanishes in the vacuum, which also gives us the
vacuum equations without going into the details of finding the
full field equations \cite{DengizTekin}. Finally the quadratic
part $ S^{(2)}_{WNMG} $, after making use of the vacuum equations
and dropping the boundary terms, reads as
\begin{equation}
\begin{aligned}
 {S}_{WNMG}^{(2)}=\int d^3x\sqrt{-\bar{g}}\bigg \{& -\frac{1}{2}(\partial_\mu \Phi^L)^2 + \Big (6 \sigma \Lambda -\frac{9 \Lambda^2}{2m^2}-
 \frac{15 \nu m^2}{2} \Big ) \Phi^2_L \\
& +\frac{2\beta +5 }{2m}(F^L_{\mu\nu})^2- \Big ( 2 \sigma m +\frac{\Lambda}{m}+\frac{m}{8} \Big ) A^2_L-\frac{1}{m}(\bar{\nabla}\cdot
A^L)^2 \\
&+\frac{1}{m}({\cal G}^L_{\mu\nu})^2-\Big(\frac{\sigma m}{2}-\frac{\Lambda}{4m} \Big )h^{\mu \nu}
 {\cal G}^{L}_{\mu \nu}-\frac{1}{8m}R^2_L \\
&+ \Big (2 \sigma \sqrt{m}+\frac{\Lambda}{m \sqrt{m}} \Big ) \Phi^L R^L - \Big ( 8 \sigma \sqrt{m}+\frac{4 \Lambda}{m
\sqrt{m}}+\frac{\sqrt{m}}{2} \Big ) \Phi^L \bar{\nabla}\cdot A^L\bigg \}.
\label{linearform}
\end{aligned}
 \end{equation}
In deriving the above expansion, in addition to the formulas in the Appendix, the following relations have
also been used
\begin{equation}
 \Phi^2= m \Big (1+2 \tau\frac{\Phi_{L}}{\sqrt{m}}+\tau^2\frac{\Phi^2_{L}}{m}
 +{\cal O}(\tau^3) \Big ), \hskip 1 cm \Big (\nabla_\mu A_\nu \Big)=\tau \bar{\nabla}_\mu A^{L}_\nu-\tau^2
\Big(\Gamma^{\gamma}_{\mu \nu} \Big)_{L} A^{L}_\gamma+{\cal
 O}(\tau^3).
\end{equation}
The first thing to observe is that to have a non-ghost and
canonically normalized (that is  $-\frac{1}{4}$) kinetic term for
the Maxwell field, we should set $\beta =- \frac{11}{4}$, which we
do from now on. As it stands, the fields are still coupled and one
should find a way to decouple them. Such a coupling between the
scalar field and the curvature is expected, since we are dealing
with a non-minimally (in fact conformally) coupled scalar field to
gravity. The scalar field also couples to the gauge field as
demanded by conformal invariance. To understand how one could
decouple these fields in (\ref{linearform}), let us study a
simpler model (scalar-tensor theory)  first, and then come back to
our problem.

\subsection{Quadratic fluctuations and the spectrum of the conformally coupled scalar-tensor theory}

We choose the 2+1 dimensional conformally coupled scalar-tensor action which is given by
\begin{equation}
 S_{S-T}=\int d^3 x \sqrt{-g} \Big (\Phi^2 R + 8 \partial_\mu \Phi \partial^\mu \Phi-\frac{\nu}{2}\Phi^6 \Big),
\label{scalartensor}
\end{equation}
and ask what the particle spectrum is around its (A)dS vacuum. By
just inspecting the action, one mistakingly namely think that the scalar
field is a ghost since it comes with a negative kinetic energy
part. But this is actually a red-herring, since the action is in
the Jordan frame one cannot draw such a conclusion from the full
non-linear theory. One must either go to the Einstein frame where
the fundamental degrees of freedom are  more transparent or, in
the Jordan frame, study the quadratic fluctuations of the fields
around the vacuum. We will do both below.

Under the conformal rescaling $g_{\mu \nu}(x)=\Omega^{-2}(x)g^E_{\mu \nu}(x)$, with
$\Omega\equiv(\frac{\Phi}{\Phi_0})^{2}$, the action (\ref{scalartensor}) transforms into the Einstein frame, as
\begin{equation}\label{steski}
^E S_{S-T}=\int d^3 x\sqrt{-g^E}
\Phi_0^{2}\Big(R^E-\frac{\nu}{2}\Phi_0^{4}\Big),
\end{equation}
in which $\Phi_0$ is a constant and introduced in order to keep
$\Omega$ dimensionless. Therefore, the conformally-coupled scalar field simply
disappears and one is left with pure cosmological Einstein theory
which has a massless spin-2 particle in its spectrum. How does one see this
result in the Jordan frame (which we need for our main problem).
We take (\ref{scalartensor}) and expand up to quadratic
order in the scalar and tensor fields about the (A)dS vacuum to get
\begin{equation}
\begin{aligned}
S_{S-T}=\int d^3x\sqrt{-\bar{g}}\bigg \{
&6m\Lambda-\frac{\nu}{2}m^3+\tau\Big[(3m\Lambda-\frac{\nu}{4})h+(12\sqrt{m}-3\nu
m^{5/2})\Phi^L+m\,R^L\Big]\\
&+\tau^2\Big[(-\frac{1}{2}m\Lambda+\frac{\nu}{8}m^3)h_{\mu\nu}^2-\frac{1}{2}m
h^{\mu\nu}{\cal
G}_{\mu\nu}^L+(\frac{1}{4}m\Lambda-\frac{\nu}{16}m^3)h^2 \\
&+2\sqrt{m}R^L\,\Phi^L
+(6\Lambda-\frac{15}{2}\nu m^2)\Phi_L^2+8(\partial_\mu
\Phi_L)(\partial^\mu\Phi_L)\Big]\bigg \}.
\end{aligned}
 \end{equation}
Again $ {\cal O}(\tau^0) $ part is not relevant. $ {\cal O}(\tau^1)$ part gives the vacuum of the theory, inserting the
value of $R_L$ (Appendix) in the linear part and dropping the
boundary terms, one obtains
\begin{equation}
\Lambda=\frac{\nu m^2}{4}.
\end{equation}
 Using this value in the quadratic part results in
\begin{equation}
S^{(2)}_{S-T}= \int d^3x\sqrt{-\bar{g}}\bigg \{-\frac{1}{2}m h^{\mu\nu}{\cal G}^L_{\mu\nu}+2\sqrt{m}R_L\Phi_L-24\Lambda\Phi_L^2
+8(\partial_\mu\Phi_L)^2 \bigg \}.
\label{quadratic}
\end{equation}
By the redefinition of tensor field as follows
\begin{equation}
h_{\mu\nu} \equiv\widetilde{h}_{\mu\nu}-\frac{4}{\sqrt{m}}
\bar{g}_{\mu\nu}\Phi_L,
\label{rede}
\end{equation}
 (\ref{quadratic}) reduces to the linearized version of the  cosmological Einstein theory
\begin{equation}
S^{(2)}_{S-T}= -\frac{1}{2}m \int d^3x\sqrt{-\bar{g}}\,\widetilde{h}^{\mu\nu}\widetilde{{\cal
G}}^L_{\mu\nu}.
\label{ste}
\end{equation}
As it is clear, just like in the Einstein frame, here at the
quadratic level of the Jordan frame, the conformally-coupled
scalar field with the wrong-sign kinetic energy disappears in
(\ref{ste}). We will use a similar field redefinition in
(\ref{linearform}).

\subsection{Weyl-invariant gauge-fixing condition}

Before we can identify the fundamental degrees of freedom, there
is one more issue that we must discuss: The gauge field
in its locally Lorentz invariant form, has spurious
(non-propagating) degrees of freedom, which we must eliminate.
This can be done with a Weyl-invariant gauge-fixing. Such a gauge condition can be found as
follows: Let the gauge-covariant derivative act on the gauge
field as \cite{DengizTekin}
\begin{equation}
\mathcal{D}_\mu A_\nu\equiv\nabla_\mu A_\nu+A_\mu A_\nu.
\end{equation}
Under the transformations (\ref{gaugetran}), in $n$ dimensions, it is easy to show that the divergence transforms as
\begin{equation}
(\mathcal{D}_\mu
A^\mu)'=e^{-2\zeta}\Big(\mathcal{D}_\mu
A^\mu-\mathcal{D}_\mu
\partial^\mu\zeta+(n-3)(A^\alpha\partial_\alpha\zeta-\partial_\alpha\zeta\partial^\alpha \zeta)\Big),
\end{equation}
so in 3 dimensions by setting $\mathcal{D}_\mu
\partial^\mu\zeta=0$, we have
 \begin{equation}
(\mathcal{D}_\mu A^\mu)'=e^{-2\zeta}(\mathcal{D}_\mu A^\mu).
\end{equation}
Therefore, we can choose a Lorenz-like condition
\begin{equation}
\mathcal{D}_\mu A^\mu=\nabla\cdot A+A^2=0, \label{gaugefixing}
\end{equation}
as a Weyl-invariant gauge-fixing condition. It is important to
note that $\mathcal{D}_\mu \partial^\mu\zeta=0$ is also
Weyl-invariant. [This is a Weyl-invariant generalization of the
leftover gauge-invariance, $ \partial^2 \zeta=0 $, after the usual
Lorenz gauge $\partial_\mu A^\mu =0$ is chosen.] At the linear
level, (\ref{gaugefixing}) reduces to the background covariant
Lorenz condition: $\bar{\nabla}\cdot A_L=0$.

These tools are sufficient to decouple the fundamental degrees of freedom in (\ref{linearform})
which we do in the next section.

\section{Particle spectrum and their masses}

The Weyl-invariant gauge-fixing term (\ref{gaugefixing}) at the
linear level eliminates the cross term between the gauge field
and scalar field in (\ref{linearform}). On the other hand,
redefinition (\ref{rede}) works well in decoupling scalar and
tensor fields, so at the end (\ref{linearform}) becomes
\begin{equation}
\begin{aligned}
\widetilde{S}_{WNMG}=\int d^3x\sqrt{-\bar{g}}\bigg
\{&-\frac{1}{2}\Big(16\sigma+\frac{8\Lambda}{m^2}+1\Big)(\partial_\mu
\Phi^L)^2 \\
&-\frac{1}{4m}(F^L_{\mu\nu})^2 -\Big(2\sigma
m+\frac{\Lambda}{m}+\frac{m}{8}\Big)(A^L_\mu)^2   \\
 &-\Big(\frac{\sigma m}{2}-\frac{\Lambda}{4m} \Big
)\widetilde{h}^{\mu \nu}
 {\cal \widetilde{G}}^{L}_{\mu \nu}
 +\frac{1}{m}({\cal\widetilde{G}}^L_{\mu\nu})^2-\frac{1}{8m}
\widetilde{R}^2_L
 \bigg \},
\label{WeylNMGde}
\end{aligned}
 \end{equation}
where we have used the following relations that arise after the
field redefinition of (\ref{rede})
\begin{equation}
\begin{aligned}
(R_{\mu\nu})_L=&(\widetilde{R}_{\mu\nu})_L+\frac{2}{\sqrt{m}}(\bar{\nabla}_\mu\partial_\nu\Phi_L+\bar{g}_{\mu\nu}\bar{\Box}\Phi_L),
\, \, \, \, \, \, \, R_L=\widetilde{R}_L+\frac{8}{\sqrt{m}}(\bar{\Box}\Phi_L+3\Lambda\Phi_L),\\
{\cal G}_{\mu\nu}^L=&\widetilde{{\cal
G}}^L_{\mu\nu}+\frac{2}{\sqrt{m}}\Big(\bar{\nabla}_\mu\partial_\nu\Phi_L-\bar{g}_{\mu\nu}\bar{\Box}\Phi_L-2\Lambda
\bar{g}_{\mu\nu}\Phi_L\Big),\\
h^{\mu\nu}{\cal G}^L_{\mu\nu}=&\widetilde{h}^{\mu\nu}{\cal
\widetilde{G}}^L_{\mu\nu}+\frac{4}{\sqrt{m}}\widetilde{R}_L\Phi_L+\frac{16}{m}\Phi_L\bar{\Box}\Phi_L+\frac{48}{m}\Lambda\Phi_L^2,\\
({\cal
G}_{\mu\nu}^L)^2=&({\cal{\widetilde{G}}}^L_{\mu\nu})^2+\frac{8}{m}(\bar{\Box}\Phi_L)^2+\frac{40}{m}\Lambda\Phi_L\bar{\Box}\Phi_L
+\frac{48}{m}\Lambda^2\Phi_L^2+\frac{2}{\sqrt{m}}\widetilde{R}_L\bar{\Box}\Phi_L+\frac{4}{\sqrt{m}}\Lambda\widetilde{R}_L\Phi_L.
\end{aligned}
\end{equation}
The expression (\ref{WeylNMGde}) is what we were looking for to
identify the fundamental excitations and their masses. The first
line shows that we have a unitary  massless scalar field as long
as we have a non-ghost kinetic term which is guaranteed by
\begin{equation}
 16\sigma+\frac{8\Lambda}{m^2}+1  \ge 0.
\label{constraint_scalar}
\end{equation}
In fact, when the bound is saturated the scalar field ceases to be
dynamical. The second line is the action for a massive spin-1
field (a Proca field) which propagates 2 unitary degrees of
freedom in 2+1 dimensions with mass-square
\begin{equation}
M_A^2= (4\sigma +\frac{1}{4})m^2+2\Lambda  \ge 0,
\end{equation}
which is exactly equal to the constraint on the kinetic energy of
the scalar field (\ref{constraint_scalar}). The third line needs a
little more explanation, since the fundamental degrees of freedom
are not transparent by a cursory look. But, the action is exactly
what one gets from the linearization of the NMG (\ref{nmg}) around
its (A)dS vacuum [albeit with fixed ratios of dimensionful
parameters]. There are two ways to find that it describes a
massive spin-2 field. The first way is to show with the help of an
auxiliary field that the action reduces to the massive Fierz-Pauli
spin-2 theory  \cite{BHT}. The second way is to explicitly
decompose the $h_{\mu \nu}$ into its irreducible components and at
the end express all the fundamental degrees of freedom as scalar
fields \cite{canonical}. These two pictures yield obviously the
same result showing that the third line of (\ref{WeylNMGde})
describes a massive spin-2 field with mass-square
\begin{equation}
M_g^2 = - \sigma m^2 + \frac{\Lambda}{2}.
\label{bf_bound}
\end{equation}
Unitarity of the massive spin-2 theory depends whether one is
dealing with an AdS ($\Lambda < 0$ ) or a dS ($\Lambda > 0$)
background. For the AdS background, Breintenlohner-Freedman (BF)
bound \cite{bf,waldron} $ M_g^2 \ge \Lambda$ must be satisfied, on
the other hand for the  dS background, Higuchi \cite{higuchi}
bound $ M_g^2 \ge \Lambda >0$ must be satisfied. What we have not
yet shown is that all the unitarity conditions on the spin-0,
spin-1 and the spin-2 fields are compatible with each other and
with the condition that the theory has a maximally symmetric
vacuum (\ref{lambda_exp}). First of all  one should notice that
$\Lambda_{+}$ corresponds to the dS and  $\Lambda_{-}$ corresponds
to the AdS spaces. It is easy to see that (\ref{bf_bound}) is not
compatible with the dS branch, therefore the theory is not unitary
in dS. On the other hand, considering all the conditions together,
one finds that the theory is unitary in AdS with a massless
spin-0, a massive spin-1 and a massive spin-2 field as long
as\footnote{Note that negative $\nu$ is not allowed for the scalar
field to have a viable potential with a lower bound. Moreover, for
$\nu =0$, the conformal symmetry cannot be broken since
Coleman-Weinberg potential at any loop would vanish in the case of
flat space, but in what follows below, we  include this point to
explore the full parameter range. }
\begin{equation}
\begin{aligned}
-&\frac{1}{16} < \sigma \le 0, \hskip 1.5 cm 0< \nu \le \frac{1}{64} ( 1- 256 \sigma^2) , \\
 &0 < \sigma \le \frac{1}{16}  , \hskip 1.5 cm 0 \le \nu \le \frac{1}{64} ( 1- 256 \sigma^2).
\end{aligned}
\end{equation}
On the other hand, for AdS  the theory has a massless spin-1
field, a massive spin-2 field (with $ M_g^2 = - \frac{ 3
m^2}{16}$) (no scalar field) for
\begin{equation}
\sigma=\frac{1}{16}, \hskip 1 cm \nu=0,  \hskip 1 cm \Lambda_{-}= -\frac{m^2}{4}.
\end{equation}
For flat vacuum the theory becomes unitary when
\begin{equation}
-\frac{1}{16} \le \sigma \le 0, \hskip 1 cm \nu=0,
\end{equation}
for which generically the theory has a massless spin-0, massive spin-1 and massive spin-2 fields.
There are two special points: for $ \sigma = -\frac{1}{16}$, there is no scalar field, there is a
massless gauge field and a massive spin-2 field with mass $M_g = \frac{m}{4}$. For $\sigma = 0$, there is
a massless spin-0, a massless spin-2 and a massive spin-1 field with $M_A = \frac{m}{2}$.

\section*{Conclusions}

By computing the action up to second order in all directions in
the space of gauge and scalar fields and  the metric, we have
shown that the Weyl-invariant extension of NMG is a unitary theory
generically describing a massive spin-2, a massive (or massless)
spin-1 and a massless spin-0 field around its AdS and flat vacua.
The mere existence of an AdS background spontaneously breaks the
conformal symmetry and provides mass to the spin-1 and spin-2
fields in analogy with the Higgs-mechanism. Breaking of the
conformal symmetry also fixes all the relevant couplings between
the fields. In flat space, dimensionful parameter (that is the
expectation value of the scalar field) comes from dimensional
transmutation in the quantum theory and the conformal symmetry is
broken at the two loop level \cite{tantekin} via the
Coleman-Weinberg mechanism. Weyl-invariant version of NMG seems to
be the only known toy model where a graviton mass is generated by
the breaking of a symmetry in such a way that the resultant mass
has a non-linear, fully covariant, local extension in terms of
quadratic curvature terms. In a separate work \cite{tanhayi}, we  study the
particle spectrum of similar Weyl-invariant quadratic theories in
generic $n$-dimensions whose actions were introduced in
\cite{DengizTekin}. It would also be interesting to add
conformally coupled spin fields to these models.

\section{\label{ackno} Acknowledgments}

The work of  B.T. is supported by the TUBITAK Grant No. 110T339. S.D. is supported by  TUBITAK Grant No. 109T748.

\section*{Appendix A: Quadratic expressions of the curvature terms}

This part compiles all the relevant tensors expanded up to second order around a background ($\bar{g}_{\mu \nu}$).
We take the expressions directly from \cite{ubinmg} for generic $n$-dimensions.
The metric perturbation $ h_{\mu \nu} $ is defined as
\begin{equation}
 g_{\mu \nu} \equiv \bar{g}_{\mu \nu}+\tau h_{\mu \nu}, \hskip 1 cm  g^{\mu \nu}=\bar{g}^{\mu \nu}-\tau h^{\mu \nu}+\tau^2 h^{\mu \rho} h^\nu_\rho+ {\cal O}(\tau^3).
\end{equation}
The Christoffel connection can be expanded as
\begin{equation}
 \Gamma^\rho_{\mu \nu}=\bar{\Gamma}{^\rho_{\mu \nu}}+ \tau \Big (\Gamma^\rho_{\mu \nu} \Big)_{L}
-\tau^2 h^\rho_\beta \Big (\Gamma^\beta_{\mu \nu} \Big)_{L} +{\cal
O}(\tau^3),
\end{equation}
from which follows the expansions of the Riemann and all the related tensors. We just need the following expressions:
\begin{equation}
 \Big (\Gamma^\rho_{\mu \nu} \Big)_{L}=\frac{1}{2}\bar{g}^{\rho \lambda} \Big (\bar{\nabla}_\mu h_{\nu \lambda}+ \bar{\nabla}_\nu
 h_{\mu \lambda}-\bar{\nabla}_\lambda h_{\mu \nu} \Big ),
\end{equation}
\begin{equation}
 \sqrt{-g} =\sqrt{-\bar{g}} \bigg[1+\frac{\tau}{2}h+\frac{\tau^2}{8} \Big (h^2-2 h^2_{\mu \nu} \Big)+{\cal O}(\tau^3) \bigg ],
\end{equation}
where  $ h=\bar{g}^{\mu\nu} h_{\mu \nu} $.
\begin{equation}
\begin{aligned}
 R{^\mu}{_{\nu \rho \sigma}}=& \bar{R}{^\mu}{_{\nu \rho \sigma}}+ \tau \Big (R{^\mu}{_{\nu \rho \sigma}} \Big)_{L}
 -\tau^2 h^\mu_\beta \Big ( R{^\beta}{_{\nu \rho \sigma}} \Big )_{L} \\
&-\tau^2 \bar{g}^{\mu \alpha} \bar{g}_{\beta \gamma} \bigg [\Big
(\Gamma^\gamma_{\rho \alpha} \Big)_{L} \Big(\Gamma^\beta_{\sigma
\nu} \Big)_{L}- \Big (\Gamma^\gamma_{\sigma \alpha} \Big)_{L}
\Big(\Gamma^\beta_{\rho \nu} \Big)_{L} \bigg ]+{\cal O}(\tau^3),
\end{aligned}
\end{equation}
where the linearized Riemann tensor is defined as follows
\begin{equation}
\Big ( R{^\mu}{_{\nu \rho \sigma}} \Big )_{L}= \frac{1}{2} \Big
(\bar{\nabla}_\rho \bar{\nabla}_\sigma h^\mu_\nu+\bar{\nabla}_\rho
\bar{\nabla}_\nu h^\mu_\sigma-\bar{\nabla}_\rho \bar{\nabla}^\mu
h_{\sigma \nu}-\bar{\nabla}_\sigma \bar{\nabla}_\rho
h^\mu_\nu-\bar{\nabla}_\sigma \bar{\nabla}_\nu
h^\mu_\rho+\bar{\nabla}_\sigma \bar{\nabla}^\mu h_{\rho \nu} \Big
). \label{riem}
\end{equation}
The quadratic expansion of the Ricci tensor follows as
\begin{equation}
\begin{aligned}
 R_{\nu \sigma}=& \bar{R}_{\nu \sigma}+\tau \Big (R_{\nu \sigma} \Big)_{L}-\tau^2 h^\mu_\beta \Big(R^\beta{_{\nu \mu \sigma}} \Big)_{L} \\
&- \tau^2 \bar{g}^{\mu \alpha} \bar{g}_{\beta \gamma} \bigg [\Big
(\Gamma^\gamma_{\mu \alpha} \Big)_{L} \Big(\Gamma^\beta_{\sigma
\nu} \Big)_{L}- \Big (\Gamma^\gamma_{\sigma \alpha} \Big)_{L}
\Big(\Gamma^\beta_{\mu \nu} \Big)_{L} \bigg ]+{\cal O}(\tau^3),
\label{ricc}
\end{aligned}
\end{equation}
where the linearized Ricci tensor is
\begin{equation}
 R^{L}_{\nu \sigma}=\frac{1}{2} \Big (\bar{\nabla}_\mu \bar{\nabla}_\sigma h^\mu_\nu+\bar{\nabla}_\mu \bar{\nabla}_\nu
 h^\mu_\sigma- \bar{\Box}h_{\sigma \nu}-\bar{\nabla}_\sigma \bar{\nabla}_\nu h \Big).
\end{equation}
The quadratic expansion of the curvature scalar is
\begin{equation}
\begin{aligned}
 R=\bar{R}+\tau R_{L}+\tau^2 \bigg \{& \bar{R}^{\rho \lambda}h_{\alpha \rho}h^\alpha_\lambda-h^{\nu \sigma}
 \Big(R_{\nu \sigma} \Big)_{L}-\bar{g}^{\nu \sigma} h^\mu_\beta \Big(R^\beta{_{\nu \mu \sigma}} \Big)_{L} \\
& -\bar{g}^{\nu \sigma} \bar{g}^{\mu \alpha} \bar{g}_{\beta
\gamma} \bigg [\Big (\Gamma^\gamma_{\mu \alpha} \Big)_{L}
\Big(\Gamma^\beta_{\sigma \nu} \Big)_{L}- \Big
(\Gamma^\gamma_{\sigma \alpha} \Big)_{L} \Big(\Gamma^\beta_{\mu
\nu} \Big)_{L} \bigg ] \bigg \}+{\cal O}(\tau^3),
\end{aligned}
\end{equation}
where
\begin{equation}\label{rl}
 R_{L}=\bar{g}^{\alpha \beta} R^{L}_{\alpha \beta}-\bar{R}^{\alpha \beta} h_{\alpha
 \beta}.
\end{equation}
The linear form of the Einstein tensor that we frequently used in
the text is
\begin{equation}
{\cal
G}_{\mu\nu}^L=(R_{\mu\nu})^L-\frac{1}{2}\bar{g}_{\mu\nu}R^L-\frac{2\Lambda}{n-2}
h_{\mu\nu}.
\end{equation}

\section*{Appendix B: Weyl-invariant action in the Einstein frame}

Here we will transform  Weyl-invariant new massive gravity (\ref{winmg}), which is necessarily in the Jordan frame to the Einstein frame.
In what follows we will keep some of the computations in $n$-dimensions for the sake of generality and set $n=3$ later. 
 $g_{\mu \nu}(x)$ denotes the Jordan frame metric and $g^E_{\mu \nu}(x)$ denotes the Einstein frame metric which are related as
\begin{equation}
 g_{\mu \nu}(x)=\Omega^{-2}(x)g^E_{\mu \nu}(x), \hskip 1.5 cm  \sqrt{-g}=\Omega^{-n}\sqrt{-g^E}.
\end{equation}
The Riemann and the Ricci tensors and the curvature scalars in the two frames are related to each other, respectively as follows 
\begin{equation}
\begin{aligned}
  {R}^\mu{_{\nu\rho\sigma}} [g]&= (R^\mu{_{\nu\rho\sigma}})^E-2
\delta^\mu{_{[\sigma}} \nabla_{\rho]} \partial_\nu\ln\Omega
-2 g_{\nu[\rho}\nabla_{\sigma]} \partial^\mu\ln\Omega \\
& \quad -2 \partial_{[\sigma}\ln\Omega \,\,  \delta^\mu_{\rho]}
\partial_\nu\ln\Omega +2 g_{\nu[\sigma} \partial_{\rho]}\ln\Omega \,\,\partial^\mu\ln\Omega  +2 g_{\nu[\rho}
\delta^\mu_{\sigma]}  (\partial_\lambda\ln\Omega)^2.
\end{aligned}
\end{equation}
\begin{equation}
\begin{aligned}
{R}_{\nu\sigma}[g]&=(R_{\nu\sigma})^E+(n-2)\Big [\nabla_\sigma
\partial_\nu\ln\Omega + \partial_\nu\ln\Omega\,\,
 \partial_\sigma\ln\Omega -g^E_{\nu\sigma}(\partial_\lambda\ln\Omega)^2  \Big ] \\
& \quad +g^E_{\nu\sigma}\Box\ln\Omega.
\end{aligned}
\end{equation}
\begin{equation}
{R}[g]=\Omega^2\Big(R^E+2(n-1)\Box\ln\Omega-(n-1)(n-2)
(\partial_\lambda\ln\Omega)^2\Big). \end{equation}
From these we can get the invariant relevant for quadratic gravity as :
\begin{equation}
\begin{aligned}
  (R_{\mu\nu\rho\sigma})^2 [g]&= \Omega^4\bigg
  \{(R^E_{\mu\nu\rho\sigma})^2+8(R^E_{\mu\nu})\nabla^\mu\partial^\nu\ln\Omega+8(R^E_{\mu\nu})\partial^\mu\ln\Omega\,\,\partial^\nu\ln\Omega-
  4R^E(\partial_\mu\ln\Omega)^2\\
  &+4(n-2)(\nabla_\mu\partial_\nu\ln\Omega)^2+4(\Box\ln\Omega)^2-8(n-2)\Box\ln\Omega(\partial_\mu\ln\Omega)^2\\
  &+8(n-2)(\partial_\mu\ln\Omega)(\partial_\nu\ln\Omega)\nabla^\mu\partial^\nu\ln\Omega
  +2(n-1)(n-2)(\partial_\mu\ln\Omega)^4\bigg \},
\end{aligned}
\end{equation}
\begin{equation}
\begin{aligned}
(R_{\mu\nu})^2[g]&=\\
& \Omega^4\bigg
  \{(R^E_{\mu\nu})^2+2(n-2)(R^E_{\mu\nu})\nabla^\nu\partial^\mu\ln\Omega+2R^E\Box\ln\Omega+2(n-2)R^E_{\mu\nu}\partial^\mu\ln\Omega\,\partial^\nu\ln\Omega\\
  &-2(n-2)R^E(\partial_\mu\ln\Omega)^2+(n-2)^2(\nabla_\nu\partial_\mu\ln\Omega)^2+(3n-4)(\Box\ln\Omega)^2\\
  & +2(n-2)^2(\partial_\mu\ln\Omega)(\partial_\nu\ln\Omega)\nabla^\mu\partial^\nu\ln\Omega-(4n-6)(n-2)\Box\ln\Omega(\partial_\mu\ln\Omega)^2\\
  &+
  (n-2)^2(n-1)(\partial_\mu\ln\Omega)^4\bigg
  \},
\end{aligned}
\end{equation}
\begin{equation}
 \begin{aligned}
R^2[g]&=\\
& \Omega^4\bigg
  \{(R^E)^2+4(n-1)R^E\Box\ln\Omega-2(n-1)(n-2)R^E(\partial_\mu\ln\Omega)^2+4(n-1)^2(\Box\ln\Omega)^2\\
  &-4(n-1)^2(n-2)\Box\ln\Omega(\partial_\mu\ln\Omega)^2+(n-1)^2(n-2)^2(\partial_\mu\ln\Omega)^4\bigg \}.
\end{aligned}
\end{equation}
As noted in the text above equation  (\ref{steski}) one should choose $\Omega\equiv(\frac{\Phi}{\Phi_0})^{2}$, where  $\Phi_0$ is a dimensionful constant 
which keeps the Einstein-frame metric dimensionless. Note that the mere requirement that there is a transformation between the Jordan and the Einstein frame 
introduces a dimensionful constant and breaks the scaling symmetry. This symmetry breaking is not the spontaneous symmetry breaking in the vacuum that we discussed in the bulk of the paper.   With the help of the above formulas, we  can now write the Einstein frame version of the  Weyl-invariant quadratic theory  (\ref{winmg}) (not to clutter the notation, below we will drop the superscript $E$ ) 
\begin{equation}
\begin{aligned}
\tilde{S}_{NMG}=\int d^3 x \sqrt{-g} \bigg \{&\sigma\Phi_0^2 \Big[R-8(\partial_\mu \ln \Phi)^2+8 A^\mu \partial_\mu \ln \Phi-2 A^2 \Big] \\
&+\Phi_0^{-2}\Big[R^2_{\mu \nu}-\frac{3}{8}R^2+4R_{\mu \nu}\nabla^\mu \partial^\nu \ln \Phi+8R_{\mu \nu}\partial^\mu \ln \Phi \partial^\nu \ln \Phi-2 R \Box \ln \Phi \\
&-2R (\nabla_\alpha \ln \Phi)^2+4(\nabla_\mu \partial_\nu \ln \Phi)^2+8(\nabla_\alpha \ln \Phi)^4-4(\Box \ln \Phi)^2 \\
&+16(\nabla_\mu \partial_\nu \ln \Phi)\partial^\mu \ln \Phi \partial^\nu \ln \Phi -8R_{\mu \nu}A^\mu \partial^\nu \ln \Phi \\
&+2R A^\mu \partial_\mu \ln \Phi-\frac{1}{2} R A^2+2R_{\mu \nu}A^\mu A^\nu+4 \Box \ln \Phi \nabla\cdot A \\
&-8 \partial_\mu \ln \Phi \partial_\nu \ln \Phi \nabla^\mu A^\nu-16 A^\mu \partial_\mu \ln \Phi (\partial_\nu \ln \Phi)^2 \\
&-4\nabla_\mu \partial_\nu \ln \Phi \Big(\nabla^\mu A^\nu+ 4A^\mu \partial^\nu \ln \Phi  \Big)+8(A^\mu \partial_\mu \ln \Phi)^2 \\
&+4 (\nabla_\mu \partial_\nu \ln \Phi)A^\mu A^\nu +4A^2 (\partial_\mu \ln \Phi)^2+(2+\beta) F^2_{\mu \nu}+(\nabla_\mu A_\nu)^2-(\nabla.A)^2 \\
&+4\nabla_\mu A_\nu \Big(A^\mu \partial^\nu \ln \Phi+ A^\nu \partial^\mu \ln \Phi \Big )-2A^\mu A^\nu \nabla_\mu A_\nu \\
&-4A^2 A^\mu \partial_\mu \ln \Phi+\frac{1}{2}A^4\Big]
-\frac{1}{2}\Phi_0^2\Big[(\partial_\mu\ln\Phi)^2+\frac{1}{4}A^2
-A^\mu\partial_\mu\ln\Phi+\nu\Phi_0^4\Big] \bigg \}.
\end{aligned}
\label{ewnmg}
\end{equation}
To simplify\footnote{We thank a very conscientious referee who not only suggested this simplification but also actully computed a different version of it  and offered several useful remarks on the rest of the paper.}  this action let us define  $\Phi = \Phi_0 e^\varphi $ and  $D_\mu \varphi \equiv \partial_\mu \varphi - \frac{1}{2} A_\mu$  which is  actually gauge invariant   since  under the gauge transformations (\ref{gaugetran})  $\varphi  \rightarrow \varphi - \frac{1}{2} \zeta(x) $.  Then  (\ref{ewnmg}) becomes

\begin{equation}
\begin{aligned}
\tilde{S}_{NMG}=&\int d^3 x \sqrt{-g} \bigg \{\Phi_0^2 \Big[ \sigma R-(8\sigma + \frac{1}{2})(D_\mu \varphi)^2 - \frac{1}{2}\nu\Phi_0^4 
+(\frac{5}{2}+\beta)\Phi_0^{- 4}  F^2_{\mu \nu} \Big] \\
&+\Phi_0^{-2}\Big[R^2_{\mu \nu}-\frac{3}{8}R^2+4R_{\mu \nu}D^\mu \varphi D^\nu \varphi  
-2R (D_\mu \varphi)^2+8(D_\alpha \varphi)^4 \\
&+16(\nabla_\mu D_\nu \varphi)\partial^\mu \varphi \partial^\nu \varphi +8 D_\mu \varphi \partial^\mu \varphi \nabla \cdot A  
-2 A^2 \nabla_\mu D_\mu \varphi    \Big]  \bigg \}.
\end{aligned}
\end{equation}

From this action, one can find the vacuum and study the excitations about it but this route, as we we noted in the text, is rather tedious compared to the Jordan frame action that we worked with. More importantly, the Einstein-frame action is not scale invariant and hence the idea, put forward above and in \cite{DengizTekin}, that graviton becomes massive after the scale symmetry gets broken spontaneously (or radiatively)  needs to be reinterpreted.

\end{document}